\documentclass{ceab} 
\usepackage[utf8]{inputenc}
\usepackage[T1]{fontenc}
\usepackage{natbib}
\usepackage[svgnames, x11names]{xcolor}
\usepackage{hyperref}
\usepackage{graphicx} 
\usepackage{url}
\usepackage{amsmath}
\usepackage{amssymb}
\usepackage{hyperref}
\usepackage{acronym}
\usepackage{enumitem}
\usepackage{setspace}

\hypersetup{
      colorlinks = true,
      linkcolor = DodgerBlue2,
      urlcolor  = NavyBlue,
      citecolor = Black,
}

\def\runningtitle{Alves Batista: ECRS2024 Proceedings}   
\def\articlenumber{X}
\def\ppages{1--21}

\acrodef{AGN}{active galactic nucleus}
\acrodefplural{AGN}{active galactic nuclei}
\acrodef{BH}{black hole}
\acrodef{BSM}{beyond the Standard Model}
\acrodef{CMB}{cosmic microwave background}
\acrodef{CR}{cosmic ray}
\acrodef{CRB}{cosmic radio background}
\acrodef{DGRB}{diffuse gamma-ray background}
\acrodef{DSA}{diffusive shock acceleration}
\acrodef{EBL}{extragalactic background light}
\acrodef{EGMF}{extragalactic magnetic field}
\acrodef{GCOS}{Global Cosmic-ray Observatory}
\acrodef{GMF}{Galactic magnetic field}
\acrodef{GRAND}{Giant Radio Array for Neutrino Detection}
\acrodef{GRB}{gamma-ray burst}
\acrodef{GZK}{Greisen-Zatsepin-Kuz'min}
\acrodef{HE}{high-energy}
\acrodef{ICS}{inverse Compton scattering}
\acrodef{IGM}{intergalactic medium}
\acrodef{IGMF}{intergalactic magnetic field}
\acrodef{ISM}{interstellar medium}
\acrodef{LHC}{Large Hadron Collider}
\acrodef{LIV}{Lorentz invariance violation}
\acrodef{QG}{quantum gravity}
\acrodef{SBG}{starburst galaxy}
\acrodefplural{SBG}{starburst galaxies}
\acrodef{SFR}{star-forming region}
\acrodef{SM}{Standard Model}
\acrodef{SMBH}{supermassive black hole}
\acrodef{SN}{supernova}
\acrodef{SNR}{supernova remnant}
\acrodef{TA}{Telescope Array}
\acrodef{TDE}{tidal disruption event}
\acrodef{UHE}{ultra-high energy}
\acrodefplural{UHE}{ultra-high energies}
\acrodef{UHECR}{ultra-high-energy cosmic ray}
\acrodef{VHE}{very-high energy}
\acrodefplural{VHE}{very-high energies}

\newcommand{\dd}{\text{d}}
\newcommand{\refsec}[1]{\S\ref{#1}}
\newcommand{\refeq}[1]{eq.~\ref{#1}}
\newcommand{\refEq}[1]{Eq.~\ref{#1}}
\newcommand{\reffig}[1]{fig.~\ref{#1}}

\newcommand{\myParagraph}[1]{\medskip \noindent $\triangleright$ \textit{#1}}

\begin{document}

\title{The Quest for the Origins of\\ Ultra-High-Energy Cosmic Rays}

\author{R. {Alves~Batista}$^{1, 2}$
\vspace{2mm}\\
\it $^1$ Sorbonne Université, Institut d'Astrophysique de Paris, \\ 
\it CNRS UMR 7095, 98 bis bd Arago 75014, Paris, France \\
\it $^2$Sorbonne Université, Laboratoire de Physique Nucléaire et de Hautes Energies \\ 
\it 4 place Jussieu, 75252, Paris, France
}

\maketitle

\begin{abstract}

Significant progress has been made over the past decades towards unveiling the sources of the most energetic particles in nature, the \acfp{UHECR}. Despite these advancements, the exact astrophysical sites capable of accelerating these particles to such extreme energies remain largely unknown. Moreover, the mechanisms by which they achieve these extreme energies are poorly understood.
Here, I provide a concise overview of the theory underlying the acceleration and propagation of \acp{UHECR}. I then critically discuss three recent results that could help unveil their origins: the reported excess around Centaurus~A, the correlation with starburst galaxies, and the efforts to jointly model the energy spectrum, composition, and arrival directions. Finally, I discuss strategies for advancing this field, emphasising the need for refined theoretical models, the challenges in building them, and the potential for new observatories to shed light on the mysteries of \acp{UHECR}.

\end{abstract}


\section{Introduction}\label{sec:intro}

\Acp{UHECR} represent one of the most remarkable and yet poorly understood phenomena in astrophysics. These particles, primarily atomic nuclei, are thought to originate from the most violent and energetic events in the universe although this issue remain open to date~\citep{alvesbatista2019d, anchordoqui2019a}. 
Their detection and interpretation are crucial for understanding the extreme environments capable of accelerating particles to energies exceeding $10^{18} \; \text{eV}$, far beyond the capabilities of terrestrial particle accelerators such as the \Ac{LHC}.

Over the past decades, ground-based \acp{UHECR} observatories such as the Pierre Auger Observatory and the \Acf{TA} have made noteworthy progress. These facilities have provided unprecedented insights into the energy spectra~\citep{ta2018a, ta2018c, auger2020c, auger2020d, auger2021i}, composition~\citep{auger2016a, auger2017c}, and arrival directions~\citep{ta2018b, auger2022d} of \acp{UHECR}, offering valuable clues about their potential sources. Nevertheless, despite these significant advancements, the fundamental questions regarding \acp{UHECR} have not been answered. \textit{What are the sources of the most energetic particle in the universe?} And \textit{how do they attain such extreme energies?}

To solve these central issues, it is important to \emph{carefully} consider how to interpret the data collected by \ac{UHECR} observatories. This task is not trivial, as it requires highly elaborate models of how \acp{UHECR} are produced, how they escape the environments surrounding their sources, and how they propagate through the universe. This, evidently, requires sophisticated computational tools for model-building, in addition to advanced statistical methods to make the most of the available data.

In this work, I  focus on the state-of-the-art methods and theoretical frameworks used to interpret \ac{UHECR} data, with emphasis on the limitations of current models and the challenges that remain. I begin in section~\refsec{sec:acceleration} by briefly presenting how \acp{UHECR} are accelerated to such extreme energies, and the key observables that are used to study them. In \refsec{sec:propagation}, I examine the propagation of \acp{UHECR} in the universe. Section~\refsec{sec:intepretation} is dedicated to a theoretical discussion of some interesting recent observations.

\section{UHECR sources and acceleration}\label{sec:acceleration}

The mechanisms whereby \acp{CR} can obtain \acp{UHE} are not fully understood. 
Conceptually, one could think of two ways for \acp{CR} to acquire their extreme energies, depending on whether they start with even higher energies and end up with the energies we detect -- the so-called \emph{top-down models} -- or whether they are accelerated to the observed energies -- the \emph{bottom-up models}.

\emph{Bottom-up models} are a more orthodox description of particle acceleration. The archetypical case is \emph{electromagnetic acceleration}, wherein electromagnetic processes in the vicinity of their sources donate energy to the \acp{CR}. This includes a variety of acceleration templates such as the well-known second-order Fermi acceleration~\citep{fermi1949a}, diffusive shock acceleration~\citep{bell1978a, bell1978b, drury1983a}, magnetic reconnection~\citep{degouveiadalpino2000a, giannios2010a}, unipolar induction~\citep{goldreich1969a, lovelace1976a}, one-shot mechanisms~\citep{caprioli2015a}, shear acceleration~\citep{berezhko1981a, berezhko1981b}, and many others~\citep{matthews2020a}.
The second type is \emph{gravitational acceleration}, wherein particles are accelerated by the gravitational field of massive objects, such as black holes~\citep{banados2009a, zaslavskii2010a, zaslavskii2012a}. Despite its low popularity and the criticisms it has received, this type of acceleration remains a valid option until it can be fully tested~\citep[see][for a review]{harada2014a}.

\emph{Top-down models} are based on the decay of supermassive particles, such as topological defects or superheavy dark matter particles, which would produce \acp{UHECR}. Intuitively, top-down models tend to create more problems than they solve, as they require an explanation for how such highly energetic particles were created in the first place.
The non-observation of \ac{UHE} photons provides evidence against several such models, imposing stringent constraints on their viability~\citep{ta2019a, auger2022c, auger2023c}.

\subsection{Some criteria for \ac{UHECR} acceleration}\label{ssec:accelerationCriteria}

Under the electromagnetic acceleration paradigm (and partly for the gravitational acceleration one), \ac{UHECR} sources generally satisfy a few constraints~\citep{ptitsyna2010a}, listed below.

\begin{figure}[htb!]
      \centering
      \includegraphics[width=0.9\columnwidth]{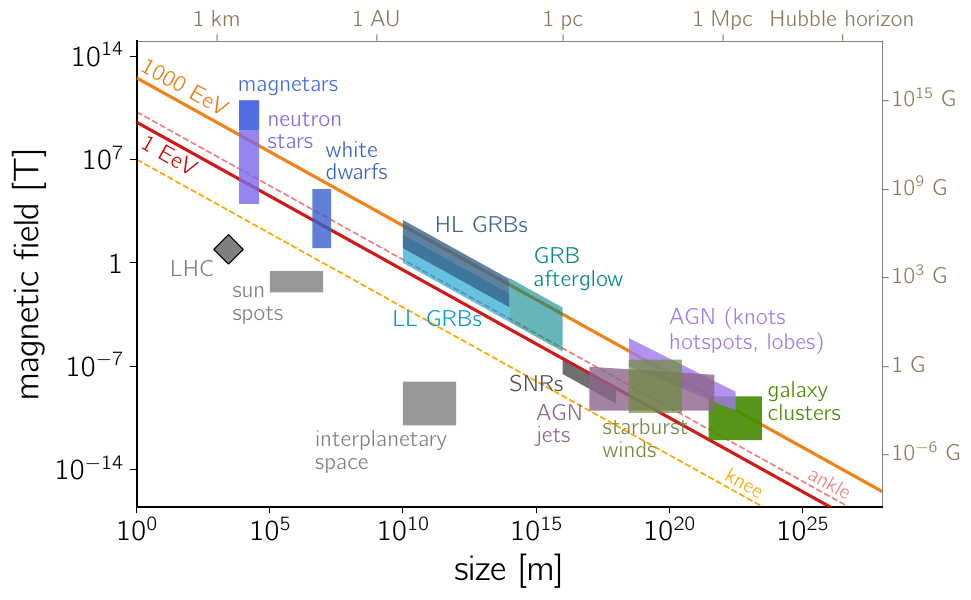}
      \caption{The \emph{Hillas diagram}, illustrating different classes of possible particle accelerators. The vertical axis represents the magnetic field strength of the confining region, whereas the horizontal refers to the size of this region, in the co-moving frame of the source. Solid lines indicate the energies above which particles can no longer be trapped by the object, under the assumption of Bohm diffusion. The dashed lines indicate the energies corresponding to the knee and ankle features of the \ac{CR} spectrum. The information used in this plot is based on reviews by \citet{alvesbatista2019d} and \citet{anchordoqui2019a}.}
      \label{fig:hillas}
\end{figure}

\myParagraph{Geometrical constraints.}
A useful geometrical criterion commonly used to identify possible \ac{UHE} accelerators was identified by \citet{hillas1984a}. The so-called \emph{Hillas condition} represents a necessary (but not sufficient) condition for particle acceleration. It states that the size of the confining region ($R_\text{o}$) must be larger than the Larmor radius of the particle ($R_\text{L}$):
\begin{equation}
	R_\text{o} \gtrsim R_\text{L} \;\; \Rightarrow \;\; R_\text{o} \gtrsim \dfrac{1}{Z} \left(\dfrac{B_\text{o}}{1 \; \text{nT}}\right)^{-1}  \left(\dfrac{E}{10 \; \text{EeV}}\right) \; \text{kpc} \,,
	\label{eq:hillas}
\end{equation}
where $B_\text{o}$ is the magnetic field in the acceleration region, $E$ is the energy of the particle, and $Z$ is the atomic number of the \ac{CR} nucleus. A schematic illustration of the Hillas criterion is shown in \reffig{fig:hillas}.
The corresponding maximum energy achievable by a given source does \emph{not} follow immediately from \refeq{eq:hillas}, as it depends on the typical velocity of the fastest particles. In this case, the geometrical condition given by \refeq{eq:hillas} leads to the following dynamical constraint on the maximal energy ($E_\text{max}$):
\begin{equation}
	E_\text{max} \lesssim Z \beta R_\text{o} B_\text{o} \;\; \Rightarrow \;\; E_\text{max} \lesssim 10 Z \beta \left(\dfrac{B_\text{o}}{1 \; \text{nT}}\right) \left(\dfrac{R_\text{o}}{1 \; \text{kpc}}\right) \; \text{EeV}  \,.
	\label{eq:hillasEmax}
\end{equation}
The proportionality of $E_\text{max}$ to $Z$ implies a rigidity dependence $R_\text{max} = E_\text{max} / Ze$. Note that, for relativistic shocks, a Lorentz factor ($\gamma$) can increase $E_\text{max}$ from \refeq{eq:hillasEmax}.
In reality, the acceleration process is much more complex than what is described by the Hillas criterion. Energy losses and magnetic field inhomogeneities can significantly affect the acceleration process. Interested readers are directed to the review by \citet{matthews2020a}. 

\myParagraph{Power constraints.}
The energy budget available for acceleration at any given instant must be sufficient to be used for acceleration. Assuming that this energy is supplied by the magnetic field, one arrives at the Hillas-Lovelace limit~\citep{lovelace1976a, hillas1984a}:
\begin{equation}
	P_{B,\text{min}} \gtrsim u_B R_\text{o}^2 v  \;\; \Rightarrow \;\; P_{B,\text{min}} \gtrsim \dfrac{10^{37} \eta^2}{\beta Z^2} u_B \left(\dfrac{E_\text{max}}{10 \; \text{EeV}}\right)^2  \; \text{J} \, \text{s}^{-1} \;,
	\label{eq:lovelace}
\end{equation}
where $\eta$ refers to an efficiency factor, and $u_B$ the magnetic energy density. \refEq{eq:lovelace} is oversimplified here and holds for the non-relativistic case. In reality, the power $P_{B,\text{min}}$ depends on details of the acceleration process and on the diffusion properties of the medium~\citep[for more detailed considerations, see, e.g.,][]{waxman1995a, matthews2020a, matthews2023a}.

\myParagraph{Energy gain vs. losses.}
The energy gained by particles during acceleration must necessarily exceed their total energy losses, including radiative losses and interactions with the surrounding medium. Information about the source environment, such as the density of matter and radiation in the medium as well as the magnetic field strength, can be effectively used to rule out potential source candidates.

\medskip
\citet{ptitsyna2010a} considered two additional conditions. The first is related to the total emissivity of a given source population, whose number density must be sufficient to account for the observed flux of \acp{UHECR}. 
While valid, this condition is weaker than the others, as it is \textit{a posteriori}, depending strongly on propagation details which are not well known (see \refsec{sec:propagation}). For example, it presumes some level of knowledge about the source population, including temporal characteristics (steady emission \textit{vs.} bursts), among other factors. The second additional condition involves multi-messenger signatures and is also \textit{a posteriori}. It requires that emissivity constraints be consistent with observational limits on the diffuse fluxes of lower-energy \acp{CR}, electromagnetic radiation, and neutrinos. This condition is stronger than the aforementioned density-emissivity limits, as it incorporates information from multiple independent probes (messengers).

\section{UHECR propagation}\label{sec:propagation}

The propagation of \acp{UHECR} through intergalactic space is a complex process that depends on the properties of the particles themselves, the magnetic fields they traverse, and the background radiation fields with which they interact (see \refsec{ssec:interactions} and \refsec{ssec:magneticFields}). Accurately modelling \ac{UHECR} propagation requires accounting for these factors, which are typically combined within computational tools designed to simulate their transport (see \refsec{ssec:propagationCodes}).

\subsection{Interactions with matter and radiation}\label{ssec:interactions}

The interaction of \acp{UHECR} with matter is generally considered negligible due to the low density of gas in the \ac{IGM}. However, it can be important in other environments~\citep{owen2018a}, playing a particularly special role in gamma-ray and neutrino emission by galaxy clusters~\citep{fang2016a, hussain2021a, condorelli2023b, hussain2023a}.

\begin{figure}[htb!]
      \centering
      \includegraphics[width=0.9\columnwidth]{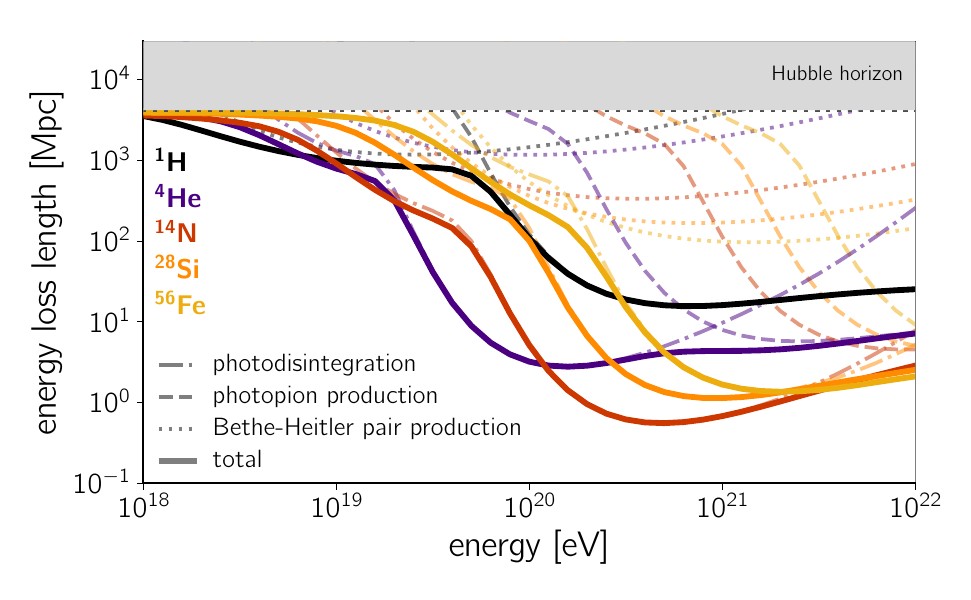}
      \caption{Energy loss length as a function of the energy, at $z=0$. Different types of nuclei are represented by distinct colours. Thin faint lines with varying styles refer to specific photonuclear process. The thick solid lines show the total contribution from all processes. Results are based on the \ac{EBL} model by \citet{saldanalopez2021a}, derived using the CRPropa code~\citep{alvesbatista2022a}.}
      \label{fig:energyLoss}
\end{figure}

One of the main interactions of \acp{UHECR} with photon fields is photomeson production -- or photopion production, since pion production is by far the dominant channel --, which is the interaction of \ac{CR} nuclei with pervasive radiation fields predominantly of the \ac{CMB}, \ac{EBL}, and \ac{CRB}. The well-known \ac{GZK} cut-off~\citep{greisen1966a, zatsepin1966a} is exactly the process of \ac{UHE} protons interacting with \ac{CMB} photons leading to a strong suppression for energies above $E_\text{GZK} \approx 5 \times 10^{19} \; \text{eV}$. \ac{UHECR} interactions with photons can incur energy losses due to the production of electron-positron pairs, a process known as Bethe-Heitler pair production~\citep{bethe1934a}. Nuclei heavier than hydrogen can also undergo photodisintegration.
Together, all these processes can significantly affect the spectrum of \acp{UHECR}. The energy loss lengths associated with these interactions, along with adiabatic energy losses due to the expansion of the universe, are illustrated in \reffig{fig:energyLoss}.

Propagation uncertainties in \acp{UHECR} were investigated by \citet{alvesbatista2015d}, who examined the effects of \ac{EBL} models and photodisintegration cross sections. Their study demonstrated that these factors can significantly influence the observed spectrum and composition. \citet{soriano2018a} conducted a detailed analysis of $^4\text{He}$ photodisintegration, revealing that its cross section had been overestimated compared to current measurements. Notably, photodisintegration cross sections also play a critical role in understanding the escape of \acp{CR} from their sources~\citep{boncioli2017a}. Additional challenges arise when considering photomeson production for heavier nuclei, as shown by \citet{morejon2019a}.

\subsection{Magnetic fields}\label{ssec:magneticFields}

\begin{figure}[htb!]
      \centering
      \includegraphics[width=0.9\columnwidth]{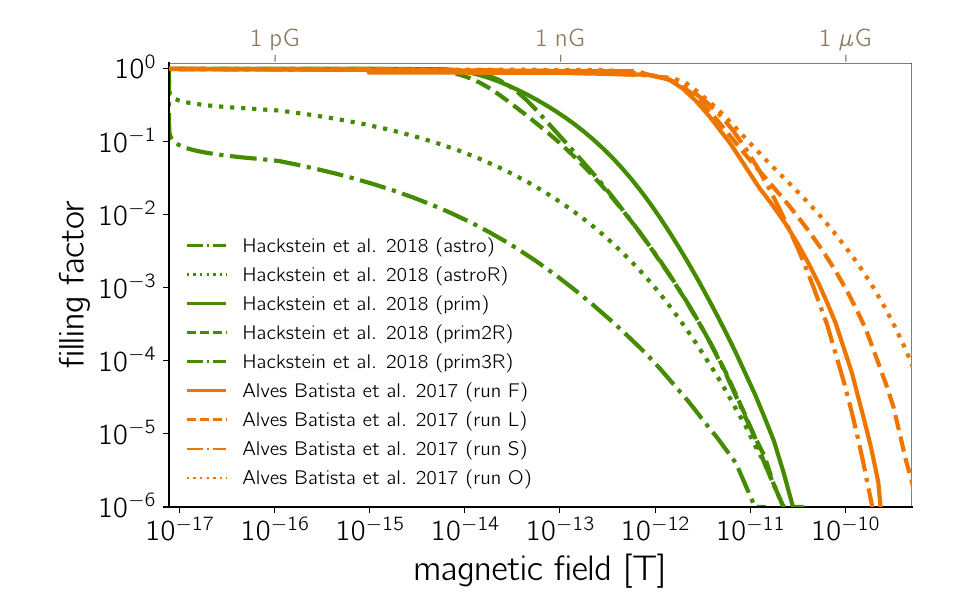}
      \caption{Cumulative volume filling factors for the \ac{EGMF} models by \citet{hackstein2018a} (green lines) and the upper-limit models of \citet{alvesbatista2017c} (orange lines) at $z = 0$. Each line style represents a different initial condition for the magnetic field, as described in the corresponding references.}
      \label{fig:egmf}
\end{figure}

The presence of \acp{EGMF} can significantly influence the propagation of \acp{UHECR}. Naturally, the \ac{EGMF} between the sources and the Earth determines by how much \acp{CR} are deflected and how much longer they must travel before reaching Earth. Since most of the universe's volume consists of ``empty'' cosmic voids, it is reasonable to expect that the fields in these regions dominate the propagation of \acp{UHECR}, although this remains to be proven. This is captured by the notion of \emph{volume filling factor}, illustrated in \reffig{fig:egmf}. One could, of course, argue that outflows from large-scale structures like galaxy clusters can inject magnetised material into the \ac{IGM}, but this effect should not significantly alter the magnetisation of cosmic voids over the age of the universe~\citep[e.g.,][]{furlanetto2001a, barai2008b, beck2013a, aramburogarcia2021a}. Therefore, unless sources are located very close to Earth such that they do not traverse cosmic voids, \acp{UHECR} would likely cross these regions at least once. 

In addition to voids, other regions of the large-scale structure of the universe also influence the propagation of \acp{UHECR}, affecting their trajectories. For instance, these structures can act as scattering centres~\citep{kotera2008b}, potentially causing significant deflections. They can also trap \acp{CR} particularly when sources are embedded within them~\citep{alvesbatista2017c, hackstein2018a}.

To a first approximation, neglecting interactions and energy losses, the deflection of \acp{UHECR} depend on the distance from Earth to the source relative to the coherence length of the field ($L_B$), yielding:
\begin{equation}
      \delta \simeq 
      \begin{cases}
            5.3^\circ Z \left(\dfrac{D}{1 \; \text{Mpc}}\right) \left(\dfrac{B_\perp}{100 \; \text{fT}}\right) \left(\dfrac{E}{10 \; \text{EeV}}\right)^{-1}  & \text{if} \; D \ll L_B \,, \\
            1.7^\circ Z \left(\dfrac{D}{1 \; \text{Mpc}}\right)^{\frac{1}{2}} \left(\dfrac{B_\perp}{100 \; \text{fT}}\right) \left(\dfrac{L_B}{100 \; \text{kpc}}\right)^{\frac{1}{2}} \left(\dfrac{E}{10 \; \text{EeV}}\right)^{-1}  & \text{if} \; D \gg L_B \,. \\
      \end{cases}
      \label{eq:deflection}
\end{equation}



The \Acf{GMF} is better understood than \acp{EGMF}, but still, much remains to be learned. New \ac{GMF} models have been proposed in the last decade~\citep{jansson2012a, jansson2012b, unger2024b, korochkin2025a}. Fortunately, in spite of the uncertainties, \acp{UHECR} can cross the galaxy without undergoing significant interactions. Consequently, their propagation can be inverted, enabling a more efficient analysis of the relevant uncertainties, using strategies such as the lensing technique~\citep{bretz2014a}.

\subsection{Propagation codes}\label{ssec:propagationCodes}

Propagation codes such as CRPropa~\citep{armengaud2007a, kampert2013a, alvesbatista2016a, alvesbatista2022a}, SimProp~\citep{aloisio2017a}, PriNCe~\citep{heinze2019a}, and TransportCR~\citep{kalashev2015a} are widely used to model the propagation of \acp{UHECR} in the universe. CRPropa and SimProp rely on Monte Carlo methods, whereas PriNCe and TransportCR employ transport equations. CRPropa also offers a three-dimensional mode, including magnetic-field effects.

None of these codes can capture the full complexity of \ac{UHECR} propagation within a feasible time frame, leaving much of the parameter space unexplored. Unfortunately, this will likely not change in the near future, particularly given the stochastic nature of some aspects of propagation, which makes any attempts to invert this problem seemingly unfeasible.
Unless we are fortunate enough that all relevant UHECR parameters lie within the small region of parameter space accessible to current methods, new techniques will be essential to advance the field. For more in-depth discussion, see the white paper of the European Consortium for Astroparticle Theory (EuCAPT)~\citep{alvesbatista2021c}.

\subsection{Cosmogenic particles}\label{sec:cosmogenic}

An important consequence of the extragalactic propagation of \acp{UHECR} is the production of \emph{cosmogenic particles}, namely neutrinos and photons. In particular, the decay of neutral ($\pi^0 \rightarrow \gamma + \gamma$) and charged ($ \pi^+ \rightarrow \mu^+ + \nu_\mu \rightarrow e^+ + \nu_e + \bar{\nu}_\mu + \nu_\mu$) pions produced via photopion production, together with beta decays ($n \rightarrow p + e^- + \bar{\nu}_e$ ) generated during photodisintegration, abundantly produce neutrinos and fuel intergalactic electromagnetic cascades initiated by photons and electrons.

An interesting aspect of cosmogenic neutrinos is that they can be used to constrain the fraction of protons in the \ac{UHECR} flux, for a given source emissivity~\citep{vanvliet2019a, muzio2023a}. However, current estimates of the flux of cosmogenic neutrinos suggest a rather bleak scenario for their detection~\citep{alvesbatista2019a, heinze2019a}. But given the limitations of the phenomenological models employed (see \refsec{ssec:combinedFit}), this pessimism might not be warranted.
Regarding the flux of cosmogenic photons, its contribution to the \acf{DGRB} at energies of $\sim 1 \; \text{TeV}$ may strongly constraint some \ac{UHECR} models~\citep{alvesbatista2019a, vanvliet2017b}.

\section{Interpreting the observations}\label{sec:intepretation}

The primary astrophysical observables derived from \ac{UHECR} data are the energy spectrum, mass composition, and arrival directions. They are essential for unraveling the origin and propagation of \acp{UHECR}, as well as for differentiating between competing source models. In the following subsections, I highlight a few relevant findings from the last decade to illustrate the state of the field, and critically examine what we have learned so far.

\subsection{Centaurus~A}\label{ssec:cenA}

One of the most compelling results from the last decade is the correlation of \acp{UHECR} and the nearest \acf{AGN}, Centaurus~A (Cen~A), located at approximately 3.8~Mpc from Earth. This radio-loud \ac{AGN} has been long deemed a strong candidate to accelerate \acp{CR} to \acp{UHE}~\citep{romero1996a}. The central galaxy itself hosts a \acl{SMBH} of approximately $55 \times 10^{6} M_\odot$, and contains a pair of radio jets, likely powered by accretion onto the central \acl{BH}, extending over several kiloparsecs~\citep{hardcastle2003a, mueller2011a, mueller2014a}. 

Several suggestions for how \acp{UHECR} could be produced in Cen~A have been put forward. The way in which particles are accelerated and the region where this happens vary hugely depending on the model, possibly taking place in the jet itself, the radio lobes, its inner core, and even in jet-lobe interactions~\citep{honda2009a, kachelriess2009a, fraija2014a, mckinley2018a, rodriguezramirez2019b}.
Cen~A is also a \ac{SBG}, displaying signs of intense star formation activity, possibly due to its history of galaxy mergers~\citep{crockett2012a}. 

The \citet{auger2022d} detected an anisotropic distribution of \acp{UHECR} at energies $E > 38 \; \text{EeV}$. A significant fraction of these \acp{CR} are clustered around the direction of Cen~A, within a $27^\circ$ window. The significance of this excess is $\approx 3.9\sigma$, depending on details of the analysis~\citep{auger2018a}. 

If the Cen~A signal continues to grow over time, by 2026 it could reach a $5\sigma$ significance. 
Suppose this is the case: \textit{could we claim that Cen~A is a source of \acp{UHECR}?} The answer is unequivocally \emph{no}. From a theoretical viewpoint, the \emph{necessary conditions} to assert that Cen~A (or any other object) is an \ac{UHECR} source are:
\begin{itemize}[noitemsep, nolistsep]
      \item The energy budget available should be sufficient to generate the observed flux of \acp{UHECR} within a given time window.
      \item If acceleration is electromagnetic, the size of the confining magnetised region must enable the escape of \acp{CR} with the energies of interest.
      \item The environment where acceleration occurs must sufficiently transparent to \acp{UHECR}, allowing their escape. Testing this \emph{necessarily} requires multi-messenger and multi-wavelength observations.
      \item The source must be unambiguously localised, meaning it has to be convincingly distinguisheable from other sources located in the vicinity or along the same line of sight.
      \item \Acp{EGMF} between the Milky Way and the object must be either reasonably well-known, to enable a precise identification of the object's coordinates in the sky, or they ought to be sufficiently small to ensure a high-confidence identification of the emitting region.
      \item The \ac{GMF} must be either sufficiently well-known to be accurately modelled, or weak enough that its impact on \acp{CR} is negligible.
\end{itemize}
These conditions are all \emph{necessary} for identifying any \ac{UHECR} source, Cen~A included, but they are \emph{not sufficient}.

\subsection{The starburst correlation}\label{ssec:starburst}

Starburst galaxies are considered potential sources of \acp{UHECR} due to their intense star formation activity, which creates an environment conducive to particle acceleration. Their magnetic fields, several orders of magnitude stronger than that of the Milky Way, enable longer confinement times, thereby allowing acceleration to higher energies~\citep{voelk1996a, bykov2014a}.
However, the highly turbulent and dense \ac{ISM} in \acp{SBG} could, in principle, absorb a significant fraction of the \acp{CR} produced within them. A way to test this hypothesis is by searching for \acl{HE} neutrinos from \acp{SBG}~\citep[e.g.,][]{loeb2006b, lunardini2019a, ambrosone2021a}. Gamma rays, which can be produced by the same mechanisms as neutrinos, might also serve as a counterpart~\citep{hess2009c, bykov2014a, ohm2016a, linden2017a}, but the \ac{ISM} could also be opaque to their propagation, making the detection of this counterpart more challenging.

The high \acl{SN} rate in \acp{SBG} likely contributes to a significantly higher flux of high-energy \acp{CR} compared to normal galaxies. Additionally, winds and outflows driven by star formation are potential accelerators~\citep[e.g,][]{anchordoqui2018b}, especially when pre-accelerated particles from \ac{SN} shocks are considered.

Currently a $4\sigma$ evidence of correlation between \acp{UHECR} and \acp{SBG} has been reported by Auger~\citep{auger2018a, auger2022d}. \textit{If this signal keeps increasing over time, can we unambiguously claim that \acp{SBG} \textit{are} sources of \acp{UHECR}?} The answer is: it depends what is meant by \emph{sources}. Given that this association is true and that there are no sources along the same line of sight, it is possible to say that \acp{UHECR} come from \acp{SBG}, but not that the \acp{SBG} themselves are sources. The necessary conditions for them to be sources are the same as for Cen~A. But there is an additional issue that deserves attention: given the high star formation rate in these galaxies, it is well possible that some objects such as young pulsars~\citep{blasi2000a, fang2012a} or magnetars~\citep{arons2003a} are the true sources.  

\citet{anchordoqui2018b} proposed that correlation with \emph{all} matter in the local universe would favour the interpretation that the sources are embedded in \acp{SBG}. He also argues that this is not viable due to energy-budget considerations, and puts forward a model based on \ac{SBG} winds.
Since Auger's results suggest that at least $\sim 10\%$ of the total \ac{UHECR} flux comes from \acp{SBG}, it is rather puzzling that \ac{TA} has not observed a significant correlation~\citep{ta2018b}.


What is more: even if we can unambiguously confirm that \acp{UHECR} come from \acp{SBG}, we are still not able to identify the acceleration mechanism. As described in \refsec{sec:acceleration}, acceleration mechanisms are simply templates and can be applied to a variety of types of objects which could also contribute to the observed flux.
Then the question becomes: \textit{what can we learn from the \ac{UHECR}--\ac{SBG} correlation?} On its own, at present, we have not learnt anything fundamentally new. However, this signal is a first step toward unveiling the sources of \acp{UHECR}. 
With higher statistics and better knowledge of propagation ingredients such as magnetic fields and background photon fields, it is possible to add information from other messengers, in particular neutrinos and gamma rays, to build a self-consistent model that describes \ac{UHECR} emission by \acp{SBG}. 
Note that multi-messenger and multi-wavelength information are \emph{necessary} conditions to meaningfully interpret these observations but they are still not \emph{sufficient}. 

In summary, to understand the apparent \ac{UHECR}--\ac{SBG} correlation, one needs to be able to disentangle between the following scenarios:
\begin{enumerate}[noitemsep, nolistsep, label = (\roman*)]
      \item astrophysical objects in \acp{SBG} that accelerate \acp{UHECR};
      \item large-scale phenomena inherent to \acp{SBG} themselves, such as super-winds, acting as  accelerators;
      \item a combination of the two, with pre-acceleration in objects and some appreciable acceleration by large-scale phenomena afterwards.
\end{enumerate}

\subsection{Combined spectrum-composition fits}\label{ssec:combinedFit}

A notable surge in the number of attempts to perform combined spectrum-composition fits has been observed in the last decade~\citep{hooper2010a, taylor2014a, taylor2015a, auger2017a, alvesbatista2019a, heinze2019a, auger2023a, auger2024c, auger2024e}. These fits are phenomenological, starting with an injection function ($Q_\text{src}$), defined as the number of particles of a given nuclear species emitted by a source, per energy, at a given time interval
\begin{equation}
      Q_\text{src} \left(E, t \right) \equiv \dfrac{\dd^2 N}{\dd E \, \dd t} = \sum\limits_{\varkappa \in \kappa} \mathcal{S}_\varkappa (E, t; E_{\text{cut}}^{(\varkappa)}(t) ) \;,
      \label{eq:injection}
\end{equation}
where $\kappa = \{ \left(A, Z\right) \; | \; \left(A > 0, \, Z \geq 0\right) \}$ is the set of all \ac{CR} species emitted, with atomic number $Z$ and atomic mass $A$. Here $E$ is the energy of the emitted \ac{CR}, $\mathcal{S}_\varkappa$ is a function that describes the spectrum, with a characteristic composition-dependent  maximal energy ($E_{\text{cut}}^{(\varkappa)}$), and $t$ is a time variable, which may be related to the redshift ($t = t(z)$) or to the intrinsic variability of the source. 

A common assumption is that \ac{UHECR} acceleration is electromagnetic and that the cut-off energy scales linearly with the particle charge ($Ze$), as argued by~\citet{peters1961a}.
Even though this assumption is reasonable and serves as a useful starting point, astrophysical environments can contain matter and radiation fields that affect $\mathcal{S}_\varkappa$~\citep{unger2015a}.

$\mathcal{S}_\varkappa$ is commonly decomposed into a power law of the form $E^{-\alpha_\varkappa}$, wherein $\alpha_\varkappa$ is the spectral index of the source, and a cut-off function $\mathcal{G}(E, t; E_{\text{cut}}^{(\varkappa)}(t))$, which is often taken as an exponential cut-off:
\begin{equation}
      S_\varkappa (E, t; E_{\text{cut}}^{(\varkappa)}(t)) = E^{-\alpha_\varkappa}\,  \mathcal{G}(E, t; E_{\text{cut}}^{(\varkappa)}(t)) \;.
\end{equation}
Recently, \citet{muzio2024a} adopted a phenomenological parametrisation for the cut-off, $E_{\text{cut}}^{(\varkappa)} = E_{\text{cut}}\left( A, Z \right) = E_0 Z^{a} A^{b}$, wherein $E_0$ is a reference energy. In this case, the Peters' cycle~\citep{peters1961a} is a special case with $a = 1$ and $b = 0$, and models with in-source photodisintegration~\citep{unger2015a} have $a = 0$ and $b = 1$. They have shown that the $\mathcal{G}$ plays an important role in the outcome of these fits. Moreover, variations of $\mathcal{G}$ across a source population can also play a role~\citep{ehlert2023a}.

\myParagraph{Source distribution.} 
A common assumption in combined fits is that the \ac{UHECR} sources are homogeneously distributed in space. This is a reasonable assumption for \acp{UHECR} of energies $E \lesssim 10 \; \text{EeV}$, whose typical energy loss lengths are comparable to or larger than the distance to the sources, such that most sources are contributing to the observed flux. However, at higher energies ($E \gtrsim 40 \; \text{EeV}$), the sources are mostly local (see \reffig{fig:energyLoss}), lying within the interaction horizon ($\lesssim 100 \; \text{Mpc}$), which means that they \emph{are not} homogeneously distributed. More recently, some progress has been made in this direction, considering specific source populations such as \acp{AGN} and minding the potential importance of possible nearby sources like Cen~A to the observed flux~\citep{auger2024c}. Moreover, new analyses looking for correlations with other messengers have also been informing on the distribution of sources~\citep{partenheimer2024a}.

\myParagraph{Source luminosity.} 
Frequently the sources are assumed to have the same luminosity, except for a few works that fix it with respect to intrinsic source properties~\citep[for instance,][]{eichmann2018a, biehl2018b, boncioli2019a, eichmann2022a}. This assumption is helpful to enable the interpretation of observations, but it is far from realistic. In reality, some sources are most likely brighter than others, depending on their size, age, and on details of the acceleration process.

\myParagraph{Composition.} 
The composition at the source is essentially the metallicity in the accelerating region, which evolves with cosmic time. Still, most works assume a time-independent composition. Although, this makes the interpretation of the data much more tractable, it is not realistic. Furthermore, the composition at the source is uncertain, and even if mean free paths are similar for a few nuclear species, the by-products of their interactions can be very different.

\myParagraph{Extragalactic magnetic fields.} 
Most works, for simplicity, ignore the effects of \acp{EGMF}. This could be justified, but at present information is lacking to support it (see \refsec{ssec:magneticFields}). \citet{wittkowski2017a} performed a combined spectrum-composition fit considering a ``realistic'' distribution of \acp{EGMF}. They found that the inclusion of \acp{EGMF} could lead to a significant change in the results of the fits. For instance, in one of their scenarios, they inferred a spectral index of $\alpha \approx 1.61$ and a maximal rigidity of $\log(R_\text{cut} / \text{eV}) \approx 18.88$, compared to $\alpha \approx 0.61$ and $\log(R_\text{cut} / \text{eV}) \approx 18.48$, in the case where \acp{EGMF} were ignored. 
More recently, the \citet{auger2024e} performed a similar fit considering the effects of turbulent \acp{EGMF} statistically (with one-dimensional simulations), which would imply a magnetic horizon effect on the \ac{UHECR} spectrum. Even though this model is completely unrealistic, considering present-day knowledge of \acp{EGMF}~\citep[for reviews, see e.g.,][]{alvesbatista2021a, vachaspati2021a}, it is a step forward compared to most studies which assume an unmagnetised universe. And most importantly, it does confirm the aforementioned trend that the best-fit spectral indices tend to increase (for an $E^{-\alpha}$ spectrum). This has important implications for understanding the transition between Galactic and extragalactic \acp{CR}~\citep{aloisio2012a, parizot2014a}, as magnetic horizon effects determine where the transition occurs~\citep[see, e.g.,][for a discussion]{alvesbatista2014b}.

\myParagraph{Temporal emission profile.} 
Phenomenological fits of the \ac{UHECR} observations rarely account for the \ac{CR} emission time. In fact, the implicit (and rather strong) assumption is that the sources are continuously emitting \acp{CR}, which is evidently incorrect. Many source candidates are transient or have some large but finite duty cycle. Moreover, even if the sources are steady emitters, their finite lifetime could have a significant impact on phenomenological observables. \citet{eichmann2023a} have shown that this could lead to a hardening of the spectrum measured from Earth, under the assumption that radio galaxies are the sources of \acp{UHECR}. Nevertheless, similar outcomes could be expected for other source populations.

\medskip
Recently, the \citet{auger2024c} extended the combined spectrum-composition fit by adding another observable: anisotropies. Nevertheless, the model does not properly account for the effects of \acp{EGMF}, simplifying it to a simple smear around the sources. \ac{UHECR} trajectories are elongated due to magnetic deflections (see \refeq{eq:deflection}), which raises interaction rates during propagation and can substantially affect the observed spectrum, composition, and anisotropies.
Previously, \citet{eichmann2018a} had already performed a combined fit of the three observables (spectrum, composition, and arrival directions), focusing on radio galaxies. 
Such multi-observable fits are essential and should become the norm, given their constraining power. Yet, they inevitably open a ``Pandora's box'' of modelling, drastically increasing the number of free parameters and hence complexity of the model.

\section{Conclusions and Outlook}\label{sec:outlook}

We find ourselves at a crossroads of \ac{UHECR} astrophysics. Existing facilities such as the Pierre Auger Observatory and the \acl{TA} are expected to continue operating for another decade, but it is unclear what will happen next~\citep{coleman2023a}. Plans are underway for new dedicated \ac{UHECR} observatories like the \acf{GCOS}~\citep{hoerandel2022a, alvesbatista2023b, gcos2025a} and POEMMA~\citep{poemma2021a}, in addition to observatories primarily designed for other messengers but which can also deliver information about \acp{UHECR}, such as the \acl{GRAND}~\citep{grand2020a}.

There is a clear need for a new generation of observatories that can provide precise measurements of the energy spectrum, mass composition, and arrival directions of \acp{UHECR}, increasing exposure and reducing uncertainties. The ability to identify mass composition on an event-by-event basis, in particular, could have a great impact on model-building~\citep[see][for recent advances in this direction]{auger2025a}.

Even if these new observatories are built, interpreting \ac{UHECR} observations is far from trivial. Major theoretical challenges persist, and the assumptions made in interpreting the data are often unjustified and at times manifestly unrealistic. 
Understandably, this can be useful for making a problem more tractable in an initial attempt to tackle it, but it should not become the default approach. Instead, models that progressively incorporate additional layers of complexity are not only more robust, but also more likely to provide an accurate representation of the \ac{UHE} universe.

Here I selected examples that illustrate why current models must be improved to provide a meaningful interpretation of the data, and how this can be done. The phenomenological fits described in \refsec{ssec:combinedFit}, for example, can be promptly enhanced with existing tools. Howerver, the vast parameter space of uncertainties still prevents a comprehensive exploration. Even so, some progress can be made without waiting for the next generation of observatories. 
The observations discussed in \refsec{ssec:cenA} and \refsec{ssec:starburst}, on the other hand, do require next-generation instruments to enable actual discoveries. 

Finally, all these efforts must be pursued within a multi-messenger framework, acknowledging the fact that no single messenger, be it \acp{CR}, photons, or neutrinos, can provide the complete picture. A synergistic approach that integrates diverse observational channels is indispensable to make progress and ultimately unveil the origins of \acp{UHECR}.

\section*{Acknowledgements} 

I acknowledge support from the Agence Nationale de la Recherche (ANR), project ANR-23-CPJ1-0103-01. I thank Michael Unger for useful comments on this manuscripts.
I also thank the organisers and participants of ECRS2024, who contributed to the discussions that inspired this work.




\footnotesize
\setstretch{0.95}


\end{document}